\documentclass[twocolumn,showpacs,preprintnumbers,amsmath,amssymb,prb]{revtex4-1}

\usepackage{graphicx}
\usepackage{dcolumn}
\usepackage{bm}

\def\rk{$R_{\mathrm{K}}$}
\def\rh{$R_{\mathrm{H}}$}
\def\rx{$R_{\mathrm{xx}}$}

\begin{document}

\preprint{}

\title{Quantum Hall effect in exfoliated graphene affected by charged impurities: metrological measurements}

\author{J. Guignard,$^{1}$ D. Leprat,$^{1}$ D. C. Glattli,$^{2}$ F. Schopfer,$^{1}$ W. Poirier$^{1}$}
\email{wilfrid.poirier@lne.fr}
\affiliation{$^{1}$Laboratoire National de M\'{e}trologie et d'Essais, 29 avenue Roger Hennequin, 78197 Trappes, France}
\affiliation{$^{2}$Service de Physique de l'\'{E}tat Condens\'{e}, Commissariat \`{a} l'\'{E}nergie Atomique, 91191 Gif-sur-Yvette, France}


\begin{abstract}
 Metrological investigations of the quantum Hall effect (QHE) completed by transport measurements at low magnetic field are carried out in a-few-$\mu\mathrm{m}$-wide Hall bars made of monolayer (ML) or bilayer (BL) exfoliated graphene transferred on $\textrm{Si/SiO}_{2}$ substrate. From the charge carrier density dependence of the conductivity and from the measurement of the quantum corrections at low magnetic field, we deduce that transport properties in these devices are mainly governed by the Coulomb interaction of carriers with a large concentration of charged impurities. In the QHE regime, at high magnetic field and low temperature ($T<1.3~\textrm{K}$), the Hall resistance is measured by comparison with a GaAs based quantum resistance standard using a cryogenic current comparator. In the low dissipation limit, it is found quantized within 5 parts in $10^{7}$ (one standard deviation, $1 \sigma$) at the expected rational fractions of the von Klitzing constant, respectively $R_{\mathrm{K}}/2$ and $R_{\mathrm{K}}/4$ in the ML and BL devices. These results constitute the most accurate QHE quantization tests to date in monolayer and bilayer exfoliated graphene. It turns out that a main limitation to the quantization accuracy, which is found well above the $10^{-9}$ accuracy usually achieved in GaAs, is the low value of the QHE breakdown current being no more than $1~\mu\mathrm{A}$. The current dependence of the longitudinal conductivity investigated in the BL Hall bar shows that dissipation occurs through quasi-elastic inter-Landau level scattering, assisted by large local electric fields. We propose that charged impurities are responsible for an enhancement of such inter-Landau level transition rate and cause small breakdown currents.
\end{abstract}

\maketitle

\section{\label{sec:level1}Introduction}

The discovery of the quantum Hall effect (QHE) in 1980\cite{Klitzing1980} has revolutionized resistance metrology by establishing a universal quantum resistance standard at rational fractions of the von Klitzing constant $R_{\mathrm{K}}\equiv h/e^2$ where $e$ is the electron charge and $h$ Planck's constant. Although the QHE was first observed in Si-MOSFETs, the cleaner two-dimensional electron gas (2DEG) made by epitaxial growth of GaAs/AlGaAs heterostructure provided more practical quantum resistance standards. They give accurate and reproducible representations of \rk~within an uncertainty below one part in $10^9$ when operated at low temperature ($T=1.5~\mathrm{K}$) and high magnetic induction ($B=10~\mathrm{T}$)\cite{Jeckelmann2001,Poirier2011}. Following observation of the QHE in graphene\cite{Novoselov2005,Zhang2005} with a sequence of Hall resistance plateaus at $R_{\mathrm{H}}=\pm R_{\mathrm{K}}/(4(n+1/2))$ (with $n$ an integer $\geqslant 0$) that survive even at room temperature\cite{Novoselov2007}, an application to resistance metrology was considered\cite{Poirier2009}. The peculiar QHE originates from the honeycomb lattice of carbon atoms in which charge carriers at low energy behave like chiral massless relativistic fermions with Berry's phase $\pi$ ~\cite{Castro2009}. Under magnetic field, the density of states becomes quantized in Landau levels (LLs) with a $4eB/h$ degeneracy (valley and spin) that occurs at energies\cite{Gusynin2005} $\pm v_{\mathrm{F}}\sqrt{2\hbar n e B}$. The robustness of the QHE on the first plateau comes from the energy spacing $36\sqrt{B[\mathrm{T}]}~\mathrm{meV}$ between the first two LLs being larger than in GaAs ($1.7B[\mathrm{T}]~\mathrm{meV}$). In bilayer graphene, which consists of two graphitic monolayers with Bernal stacking, the dispersion relation becomes parabolic and carriers behave like chiral massive ($m=0.033\times m_{e}$ with $m_{e}$ the electron mass)\cite{Ando} Dirac fermions with Berry's phase $2\pi$ ~\cite{Castro2009}. This leads to QHE\cite{Novoselov2005} with resistance plateaus at $R_{\mathrm{H}}=\pm R_{\mathrm{K}}/(4(n+1))$, with $n$ an integer $\geqslant 0$. The energy gap between LLs occuring at\cite{McCann2006} $\pm \hbar \omega_{\textrm{c}}\sqrt{n(n-1)}$ ~($\omega_{\textrm{c}}=eB/m$ is the cyclotron pulsation) is smaller than in single graphene layer, especially at low magnetic field, but is larger than in GaAs systems. Larger energy gaps give much hope that a more practical resistance standard operating at a lower magnetic field or a higher temperature could be developed in both graphene systems. In the short term, comparison of the Hall resistance in graphene systems and in GaAs would constitute a stringent test of the QHE universality. This would support ongoing efforts to make an historic evolution towards a \emph{Syst\`{e}me International} of units directly linked to fundamental constants of physics\cite{Poirier2010}. More generally, the metrological approach can supplement the understanding of physics to the limits of instrumentation. Lastly, meeting the very demanding metrological requirements for the QHE application in graphene (quality of electrical contacts, control of electronic properties such as mobility and density over large mm-size scale) further enhances the severeness of the benchmark test offered by the QHE for the quality of any two-dimensional material, and makes it very significant and useful for the development of industrial applications such as microelectronics. The metrological investigation has started shortly after the discovery of the QHE in graphene. Previously, the Hall resistance \rh~was demonstrated to agree with \rk/2 on the plateau corresponding to Landau level filling factor $\nu=n_{\mathrm{s}}h/(eB)=2$ in exfoliated monolayer graphene within a relative uncertainty of 15 parts in $10^{6}$ (one standard deviation, $1\sigma$), probably limited by the high resistance of contacts ($>1~\mathrm{k\Omega}$)\cite{Giesberg2009}. More recently, Tzalenchuk and co-workers have reported an agreement within an uncertainty as low as $9$ parts in $10^{11}$ ($1\sigma$) in a large sample ($160\times 35~\mu \mathrm{m}^2$) made of epitaxial monolayer graphene grown on the Si-terminated face of silicon carbide (SiC), with a mobility of about $7500~\mathrm{cm}^{2}V^{-1}s^{-1}$ when placed at $B=14~\textrm{T}$ and $T=0.3~\textrm{K}$ ~\cite{Tzalenchuk2010, Janssen2011}. Achieving the QHE quantization in graphene with similar uncertainty at a few teslas magnetic induction and higher temperature, which is required to develop a quantum resistance standard challenging the GaAs ones, is still a critical issue.

In this paper, we report on the accurate investigation of the QHE quantization in monolayer and bilayer exfoliated graphene lying on $\textrm{Si/SiO}_{2}$ substrate. Measurements were performed with a Cryogenic Current Comparator (CCC)-based resistance bridge. The objective was to determine limitations to the quantized Hall resistance accuracy that can be experienced in exfoliated graphene, which however turned out to be the reference technique enabling to unveil most of chiral Dirac fermions electronic transport properties. The understanding of these limitations could even be useful to overcome likely obstacles in the development of quantum resistance standards with higher performances in graphene grown either on SiC or by Chemical Vapor Deposition (CVD).

The paper is organized as follows. In Section II, we report on electronic transport properties of graphene investigated by means of conductivity measurements at low magnetic field. In both the ML and BL samples, the analysis of the conductivity dependence on charge carrier density shows that carriers are mainly scattered by a large concentration of charged impurities located about $1~\textrm{nm}$ close to the graphene flakes. The major impact of charged impurities responsible for strong spatial fluctuations of the carrier density which survive at finite density is also confirmed by measurements of quantum corrections to conductance (weak localization and universal conductance fluctuations) in the BL sample. Section III reports on quantization tests performed by means of comparing the QHE in GaAs and in graphene systems. For monolayer and bilayer graphene, the Hall resistance of the first plateau (Landau levels are spin and valley degenerated) in the zero dissipation limit is found quantized within 5 parts in $10^7$ ($1\sigma$) to \rk/2 and \rk/4 respectively. One main limitation to accuracy is the low value of the QHE breakdown current limited to about $\sim 1 ~\mathrm{\mu A}$. In section IV, we show that the mechanism of dissipation (or backscattering) in the BL sample, which ends up in the QHE breakdown is based on quasi-elastic inter-Landau level scattering (QUILLS) assisted by large local electric fields. This leads to discussing the role of charged impurities in enhancing inter-Landau level transitions.

\begin{figure}[h]
\begin{center}
\includegraphics[width=8.5cm]{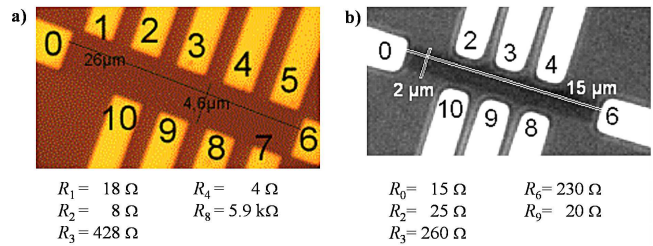}
\caption{(Color online) Optical images of the BL sample a) and of the ML sample b) with contacts resistance values indicated below.}\label{fig1}
\end{center}
\end{figure}

\section{\label{sec:level2}Electronic transport properties at low magnetic field}

Measurements were carried out on $15\times 2~\mu \textrm{m}^{2}$ and $26\times 4.6~\mu \textrm{m}^{2}$ Hall bars based on monolayer graphene (ML) and bilayer graphene (BL) respectively, which have been mechanically exfoliated from natural graphite (see Fig.~\ref{fig1}). Flakes were transferred on top of highly doped silicon substrates covered by $90~\textrm{nm}$ (resp. $500~\textrm{nm}$ in BL) of thermally grown $\textrm{SiO}_{2}$ used for backgating. Graphene flakes are electrically contacted using Ti/Au (BL) and Pd (ML) pads. Samples were then patterned with a Hall geometry appropriate for QHE precision measurements.
 Graphene arms, at least $300~\textrm{nm}$ long, connect voltage metallic contacts to the main channel. This geometry also avoids electrode-induced doping  of the main channel. Samples were finally covered with a $300~\textrm{nm}$-thick polymethyl methacrylate (PMMA) resist layer. Transport properties were explored by four-terminal resistance measurements defined by $R_{\mathrm{ij,kl}}=(V_{\mathrm{k}}-V_{\mathrm{l}})/I_{\mathrm{i} \rightarrow \mathrm{j}}$, where $V_{\mathrm{i}}$ is the voltage potential at terminal $\mathrm{i}$ and $I_{\mathrm{i} \rightarrow \mathrm{j}}$ is the current flowing between terminals $\mathrm{i}$ and $\mathrm{j}$.

\subsection{\label{sec:level2} Influence of charged impurity scattering on conductivity}

\begin{figure}[b]
\begin{center}
\includegraphics[width=8.5cm]{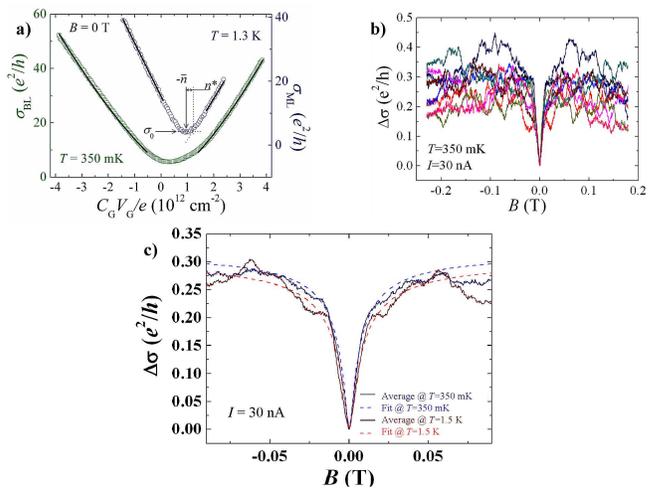}
\caption{(Color online) a) Conductivity as a function of carrier density controlled by the back-gate voltage for the BL (green) and ML (blue) samples. Solid lines are fits given by a Boltzmann transport theory including charged impurities. b) Magneto-conductivity in the BL sample at $T=0.35~\mathrm{K}$ at carrier densities in the range $n_{\mathrm{s}}=-2\times 10^{12}~\mathrm{cm^{-2}}\pm \Delta n_{\mathrm{s}}/2$ with $\Delta n_{\mathrm{s}}=3.3\times 10^{11}~\mathrm{cm}^{-2}$. c) Magneto-conductivity after averaging on carrier density and adjustments by an appropriate weak-localization theory (dotted lines) at $T=0.35~\mathrm{K}$ (blue) and $T=1.5~\mathrm{K}$ (red).}\label{fig2}
\end{center}
\end{figure}

In both samples, the four-terminal conductivity $\sigma = 1/\rho= 1/R_{\mathrm{ij,kl}}\times \frac{d_{\mathrm{kl}}}{W}$ ($W$ is the sample channel width, $d_{\mathrm{kl}}$ the distance between terminals $\mathrm{k}$ and $\mathrm{l}$), deduced from $R_{06,23}$ and $R_{18,24}$ measurements in the ML and BL samples respectively, was analyzed at zero magnetic field as a function of the gate voltage $V_{\mathrm{G}}$. It shows a typical minimum that occurs at $V_{\mathrm{Gmin}}$ (see Fig.~\ref{fig2}a). At this value the carrier density defined as $n_{\mathrm{s}}=C_{\mathrm{G}}(V_{\mathrm{G}}-V_{\mathrm{Gmin}})/e$ (with $C_{\mathrm{G}}/e=2.40\times10^{11}~\textrm{cm}^{-2}/\textrm{V}$ for ML and $C_{\mathrm{G}}/e=4.31\times10^{10}~\textrm{cm}^{-2}/\textrm{V}$ for BL) is zero on spatial average.
$\bar{n}=-C_{\mathrm{G}}V_{\mathrm{Gmin}}/e$ is the carrier density induced in the graphene by surrounding charged impurities.
While annealing the samples under vacuum at a temperature of about $400 ~\mathrm{K}$, the conductivity dip becomes sharper and its
position $V_{\mathrm{Gmin}}$ shifts near zero indicating an increase in the carrier mobility $\mu$ and a decrease of $\vert \bar{n}\vert$.
At $T=1.3 ~\mathrm{K}$ for the ML sample (resp. $T=0.35 ~\mathrm{K}$ for the BL sample) and at carrier density away from the region of the minimum conductivity, $\sigma(V_{\mathrm{G}})$ is quite linear for ML with no proof of sublinearity in the considered density range ($<2.5\times10^{12}~\textrm{cm}^{-2}$), and is slightly superlinear for BL. These features indicate that the long-range Coulomb potential induced by charged impurities constitute the dominant source of scattering in the considered samples \cite{Adam2007PRL,Adam2008}.

Conductivity for the ML sample, except near the minimum ($-\bar{n}\pm 5\times 10^{11}~\mathrm{cm}^{-2}$), is well fitted by the theoretical model based on Boltzmann transport theory with charged scatterers\cite{Adam2007,Hwang2008} $\sigma (C_{\mathrm{G}}V_{\mathrm{G}}/e)=\sigma (n_{\mathrm{s}}-\bar{n})= G(r_{\mathrm{s}},d)\frac{e^{2}}{h}\frac{\vert n_{\mathrm{s}}\vert}{n_{\mathrm{i}}}$ valid for electrons ($n_{\mathrm{s}}>n^{\ast}$) and for holes ($n_{\mathrm{s}}<-n^{\ast}$) where $n_{\mathrm{i}}$ is the density of charged impurities at an average distance $d$ from the conductor (in the silicon substrate or in the PMMA). $n^{\ast}$ is a residual density corresponding to the density of electron and hole puddles into which the system breaks at low density because of the inhomogeneous density profile created by Coulomb impurities. $r_{\mathrm{s}}$ describes the full dielectric environment of the sample that screens Coulomb interactions. Considering two semi-infinite media made of $\textrm{SiO}_{2}$ and PMMA on top of the device with dielectric constants $\epsilon_{\textrm{SiO}_{2}}=3.9$ and  $\epsilon_{\textrm{PMMA}}=4.5$
respectively, $r_{\mathrm{s}}=2e^{2}/(4\pi \epsilon_{0}(\epsilon_{\textrm{PMMA}}+\epsilon_{\textrm{SiO}_{2}})\hbar v_{\textrm{F}})=0.47$ (with $v_{\textrm{F}}=1.1\times 10^{6}~\textrm{ms}^{-1}$ ~\cite{Martin2009}) and $G(r_{\mathrm{s}}=0.47,d=0)=28.2$ in the random phase approximation. It appears that the dielectric constant of PMMA higher than air or vacuum screens more efficiently the coulombic potential of charged impurities.
Note that since $G(r_{\mathrm{s}},d)$ is only weakly dependent\cite{Adam2007PRL} on $d$, the approximated value $G(r_{\mathrm{s}},d=0)$ is valid while the electron/hole asymmetry in the conductivity curve remains weak, as observed, and thus is not considered. The mean impurity density (electron/hole average) deduced from the adjustment is $n_{\mathrm{i}}\approx1.9\times 10^{12}~\mathrm{cm}^{-2}$.
At low density, assuming this value of $n_{\mathrm{i}}$ and a finite value of $d$ in the range lower than $2~\mathrm{nm}$, the Boltzmann transport theory \cite{Adam2007} correctly explains (within a factor of 2, see ref.\cite{McCaughan}) the experimental values of the conductivity minimum $\sigma_{0}$, of the plateau width minimum conductivity $n^{\ast}$ and of the minimum position $-\bar{n}$. The size $\xi$ and the density $n^{\ast}$ of electron/hole puddles near the charge neutrality point (CNP) can be calculated from\cite{Shklovskii2007} $\xi = 1/(r_{\mathrm{s}}^{2}\sqrt{n_{\mathrm{i}}})=32~\textrm{nm}$ and $n^{\ast}=\sigma_{0}n_{\mathrm{i}}/G(r_{\mathrm{s}}=0.47,d=0)\frac{h}{e^{2}}=2.7\times 10^{11}~\mathrm{cm}^{-2}$ (with $\sigma_{0}=4\frac{e^{2}}{h}$) respectively. One deduces that each puddle contains about $9$ elementary charges in average.
The theoretical model can also explain the conductivity curve asymmetry which corresponds to a constant mobility ($\mu = \sigma /(n_{\mathrm{s}}e)$) higher for holes ($4050~\mathrm{cm^{2}\mathrm{V}^{-1}\mathrm{s}^{-1}}$) than for electrons ($3400~\mathrm{cm^{2}\mathrm{V}^{-1}\mathrm{s}^{-1}}$) by a typical $<+5~\textrm{\AA}$-size shift of the distance $d$ of charged impurities from the graphene layer under the electric field effect produced by the back-gate voltage, assuming unequal numbers of random positively and negatively charged impurities\cite{Adam2007PRL}. A similar electron/hole asymmetry has already been observed in dirty samples\cite{Tan2007,Ponomarenko}. On the other hand this asymmetry cannot be explained by the theory for attractive \textit{vs.} repulsive scattering of massless Dirac fermions by Coulomb impurities\cite{Novikov2007} predicting a higher mobility for electrons for a negative value of $\bar n$. Neither can a local doping due to the presence of metallic contacts on graphene account for it since they are non-invasive in the studied samples\cite{Huard2008} and would have induced sublinearity of the $\sigma(C_{\mathrm{G}}V_{\mathrm{G}}/e)$ curve.

In the BL sample, the conductivity can also be well fitted by a similar transport theory\cite{Adam2007PRB} based on Coulomb interactions with charged impurities $\sigma (C_{\mathrm{G}}V_{\mathrm{G}}/e)=\sigma (n_{\mathrm{s}}-\bar{n})\approx\frac{16}{\pi}\frac{e^{2}}{h}\frac{\vert n_{\mathrm{s}}\vert}{n_{\mathrm{i}}}[1+\frac{1216}{105\pi}\sqrt{\vert n_{\mathrm{s}}\vert}(d+q_{\textrm{TF}}^{-1})]$ valid for electrons ($n_{\mathrm{s}}\gtrsim n^{\ast}$) and holes ($n_{\mathrm{s}}\lesssim -n^{\ast}$) with ${q_{\textrm{TF}}}^{-1}=4\pi\epsilon_{0}(\epsilon_{\textrm{PMMA}}+\epsilon_{\textrm{SiO}_{2}})\hbar^{2}/8me^{2}=0.6 ~\textrm{nm}$ the Thomas-Fermi screening length. This theory therefore includes the superlinearity which results in a mobility depending on carrier density. Not only does the conductivity adjustment give $n_{\mathrm{i}}\approx 2\times 10^{12}~\mathrm{cm}^{-2}$ but also $d\approx 1~\textrm{nm}$. These values are comparable to values extracted from the ML conductivity curve. Taking into account that the two samples have been made using the same technological processes and the same substrates (except for the $\textrm{Si/SiO}_{2}$ thickness), this agreement strongly supports our description of conductivity using the Boltzmann transport theory based on long-range Coulomb scatterers. The extracted distance $d$ is consistent with the position of charged impurities assessed in the ML sample and generally measured in $\textrm{Si/SiO}_{2}$ substrate\cite{McCaughan}. It appears that this model correctly predicts, except for the minimum position ($-\bar{n}$), the experimental values of $\sigma_{0}$ and of the plateau width minimum conductivity $n^{\ast}$. From the values of $\xi=11~\textrm{nm}$ calculated with the specific model developed for BL \cite{CommentBLG} and the value of $n^{\ast}=\sigma_{0}\frac{\pi}{16}\frac{h}{e^{2}}n_{\mathrm{i}}=2.2\times 10^{12}~\mathrm{cm}^{-2}$ (with $\sigma_{0}=5.5\frac{e^{2}}{h}$), one deduces that each puddle contains about $8$ elementary charges.
An electron/hole asymmetry of the conductivity is also observed. But contrary to the ML sample, at $n_{\mathrm{s}}=2\times 10^{12}~\mathrm{cm}^{-2}$ the electron mobility ($2300~\mathrm{cm}^{2}\mathrm{V}^{-1}\mathrm{s}^{-1}$) is higher than the hole mobility ($2000~\mathrm{cm}^{2}\mathrm{V}^{-1}\mathrm{s}^{-1}$) by about $15\%$. It can again be explained by the shift of the mean distance $d$ between the charged impurities and the graphene layer by a few $\textrm{\AA}$ under the electric field produced by the voltage on the back-gate, but with impurities in excess with a sign opposite to the ML sample case\cite{Adam2007PRL}. We note that the same amount of charged impurities leads to a lower carrier mobility in the BL sample than in the ML sample, confirming that long-range Coulomb scattering is a very efficient mechanism to spoil mobility in bilayer graphene.
Beyond providing a very efficient source of scattering, charged impurities give rise to strong spatial fluctuations of the charge carrier density with a correlation length $\xi$ of $32~\textrm{nm}$ and $11~\textrm{nm}$ for ML and BL samples respectively. These fluctuations leading to electron/hole puddles landscape near the CNP are also known to persist at the higher carrier densities ($n_{\mathrm{s}}\approx 1-2 \times 10^{12}~\mathrm{cm}^{-2}$) where the QHE has been investigated\cite{Rossi2008,Sarma2010}. Also, a more macroscopic inhomogeneity of the carrier density at a $\mu\textrm{m}$-size scale with a typical amplitude of a few $ 10^{11}\mathrm{cm}^{-2}$ has been observed. In particular, it manifests itself in the BL sample through spatial variation of $-\bar{n}$, $\sigma_{0}$ and of $\mathrm{d}\sigma/\mathrm{d} V_{\mathrm{G}}$ slopes. These quantities depend for instance on the conductor area probed in different configurations (\textit{e.g.} $R_{18,24}$ and $R_{17,34}$). These carrier density fluctuations can be explained by spatial variations of $n_{\mathrm{i}}$ by a few $10^{11}~\mathrm{cm^{-2}}$ and of $d$ by a few $\textrm{\AA}$, which is also the typical height of graphene flake ripples. Therefore the samples are far from being homogeneous compared to GaAs based 2DEG commonly used for quantum resistance standards, where less than $10^{10}~\mathrm{cm^{-2}}$ variation of $n_{\mathrm{s}}$ can be achieved.

In both samples, the diffusion coefficient $D$ and transport mean free path $l_{\textrm{tr}}$ can be determined from conductivity measurements at low temperature and carrier densities where the QHE was investigated. In the ML sample, at $T=1.3 ~\mathrm{K}$, and $n_{\mathrm{s}}= 6.4 \times 10^{11}~\mathrm{cm}^{-2}$ (electrons), corresponding to a Fermi energy of $E_{\textrm{F}}=\hbar v_{\textrm{F}}\sqrt{n_{\mathrm{s}}\pi}=102 ~\textrm{meV}$, $D$ is calculated using the Einstein relation $D =\sigma(n_{\mathrm{s}})\sqrt{\pi}\hbar v_{\textrm{F}}/(2e^{2}\sqrt{n_{\mathrm{s}}})=2.0\times 10^{-2}~\mathrm{m}^{2}\mathrm{s}^{-1}$ and then $l_{\textrm{tr}}=2D/v_{\textrm{F}}=36~\textrm{nm}$. In the BL sample, at $T=0.35 ~\mathrm{K}$, and $n_{\mathrm{s}}=-2 \times 10^{12}~\mathrm{cm}^{-2}$ (holes), \textit{i.e.} $E_{\textrm{F}}=\hbar \sqrt{n_{\mathrm{s}}\pi}/m =72 ~\textrm{meV}$, $D=\sigma(n_{\mathrm{s}})\pi\hbar^{2}/(2e^{2}m)=1.5\times 10^{-2}~\mathrm{m}^{2}\mathrm{s}^{-1}$, and $l_{\textrm{tr}}=2D/v_{\textrm{F}}(n_{\mathrm{s}})=34~\textrm{nm}$, with $v_{\textrm{F}}(n_{\mathrm{s}})=\hbar \sqrt{n_{\mathrm{s}}\pi}/m=8.8\times 10^{5}~\mathrm{m}\mathrm{s}^{-1}$. These values confirm that electronic transport is diffusive with similar amount of disorder in both samples: $k_{\textrm{F}}l_{\textrm{tr}}=5.2$ for ML and $8.5$ for BL. Comparatively, in cleaner GaAs/AlGaAs Hall bars used as quantum resistance standards, with typical mobilities $280 000~\mathrm{cm}^{2}\mathrm{V}^{-1}\mathrm{s}^{-1}$ and density $5.2 \times 10^{11}~\mathrm{cm}^{-2}$, $l_{\textrm{tr}}=3.4~\mu\textrm{m}$ is $100$ times higher and $k_{\textrm{F}}l_{\textrm{tr}}\approx 600$.

 \subsection{\label{sec:level2} Quantum corrections to conductivity}

In the BL sample, where the QHE has been more extensively studied, quantum interference corrections to conductivity, both weak-localization correction (WL) and reproducible mesoscopic conductance fluctuations (CF), were investigated by performing magneto-conductivity measurements at low temperature. Actually, both in monolayer and bilayer graphene the amplitude of these corrections is not only ruled not by inelastic scattering like in any other diffusive metal, but also by elastic scattering mechanisms affecting the valley symmetry (intravalley scattering and/or trigonal warping of the conical band structure, intervalley scattering). This is a consequence of the direct manifestation of chirality property in quantum interference effects. For instance, interferences between time-reversal symmetric diffusive electron trajectories lead to weak-localization\cite{Kedchedzi2007} corrections to conductivity in bilayer graphene because of the charge carrier wave function $2\pi$ Berry's phase while weak-antilocalization\cite{McCannWL2006} is expected in monolayer due to $\pi$ Berry's phase.

Applying a magnetic field breaks the system time reversal symmetry and suppresses the weak-localization corrections. This gives rise to a well-known magnetoresistance.
Four-terminal magnetoresistance measurements $R(B)$ were carried out at temperatures $T=0.35~\mathrm{K}$ and $T=1.5~\mathrm{K}$, at low magnetic field using a standard AC low-frequency ($13~\textrm{Hz}$) lock-in technique, and a low-noise preamplifier. The measurement current is $I=30~\textrm{nA}$, thus the effective temperature of carriers assessed by $T_{\mathrm{eff}}=eRI/k_{\textrm{B}}=0.52~\mathrm{K}$, where $R$ is the resistance per square, is slightly higher than the base temperature $0.35~\mathrm{K}$.
Fig.~\ref{fig2}b reports a set of magnetoconductivity curves recorded at densities around $n_{\mathrm{s}}=-2 \times 10^{12}~\mathrm{cm}^{-2}$ over a total range $\Delta n_{\mathrm{s}}=3.3\times 10^{11}~\mathrm{cm}^{-2}$ . They all display a characteristic dip at zero field, signature of the expected weak-localization, the amplitude of which barely exceeds reproducible fluctuations (CF) which are analyzed below. To make the WL conductivity dip stand out from fluctuations, magnetoconductivity curves were averaged over the full density range where the diffusion coefficient $D$ does not vary by more than $10\%$. The averaged curve (see Fig.~\ref{fig2}c) is then adjusted by the appropriate weak-localization theory \cite{Kedchedzi2007}, $\Delta\sigma(B)=\sigma(B)-\sigma(0)=\frac{e^{2}}{\pi h}[F(\frac{\tau_{B}^{-1}}{\tau_{\Phi\textrm{WL}}^{-1}})-F(\frac{\tau_{B}^{-1}}{\tau_{\Phi\textrm{WL}}^{-1}+2\tau_{\mathrm{i}}^{-1}})+2F(\frac{\tau_{B}^{-1}}{\tau_{\Phi\textrm{WL}}^{-1}+\tau_{\mathrm{i}}^{-1}+\tau_{\ast}^{-1}})]$.
Here $F(z)=ln(z)+\psi(1/2+z^{-1})$, $\psi(x)$ is the digamma function, $\tau_{B}^{-1}=4eDB/\hbar$. $\tau_{\Phi\textrm{WL}}^{-1}=D/L_{\Phi\textrm{WL}}^{2}$ is the phase breaking rate. $\tau_{\mathrm{i}}^{-1}=D/L_{\mathrm{i}}^{2}$ is the intervalley scattering rate lifting the valley degeneracy of electronic states and which is caused by short-range defects with maximum size of the order of the lattice spacing. $\tau_{\ast}^{-1}=D/L_{\ast}^{2}=2\tau_{\mathrm{z}}^{-1}+\tau_{\mathrm{w}}^{-1}$ is an intravalley scattering rate. $\tau_{\mathrm{z}}^{-1}$ is the intravalley chirality breaking rate caused by surface ripples, dislocations and atomically sharp defects,\textit{i.e.} short-range defects. $\tau_{\mathrm{w}}^{-1}$ is the intravalley $p\to -p$ symmetry breaking rate (where $p=\hbar k_{\textrm{F}}$, $k_{\textrm{F}}$ is the carrier momentum at the Fermi level) caused by the anisotropy of the Fermi surface in $k$ space, \textit{i.e.} the trigonal warping. In bilayer graphene, assuming a quadratic Hamiltonian, it is expected that $\tau_{\mathrm{w}}^{-1}=\tau_{\mathrm{tr}}^{-1}$ where $\tau_{\mathrm{tr}}=l_{\textrm{tr}}/v_{\textrm{F}}$ is the transport time \cite{Kedchedzi2007}.

The adjustments of data at $T=0.35~\mathrm{K}$ and $T=1.5~\mathrm{K}$ give the phase coherence length $L_{\Phi\textrm{WL}}(T=0.35~\mathrm{K})=0.47~\mathrm{\mu m}$ and $L_{\Phi\textrm{WL}}(T=1.5~\mathrm{K})=0.42~\mathrm{\mu m}$, the intervalley scattering length $L_{\mathrm{i}}(T=0.35~\mathrm{K})=0.51~\mathrm{\mu m}$ and $L_{\mathrm{i}}(T=1.5~\mathrm{K})=0.47~\mathrm{\mu m}$, the intravalley scattering length $L_{\ast}(T=0.35~\mathrm{K})$ and $L_{\ast}(T=1.5~\mathrm{K})$ $\lesssim 0.03~\mathrm{\mu m}$. The extracted values are very similar to those measured in bilayer graphene and reported in the literature\cite{Gorbachev2007}. $L_{\Phi\textrm{WL}}$ below the sample size indicates that electronic transport is not fully quantum coherent.
It appears that $L_{\Phi\textrm{WL}} \thicksim L_{\mathrm{i}}$ and $L_{\mathrm{i}}\gg L_{\ast}$. The WL is made observable due to significant intervalley scattering, though much less than intravalley processes. It also appears that $L_{\ast}\thicksim l_{\textrm{tr}}$. The fact that $\tau_{\ast}^{-1}=2\tau_{\mathrm{z}}^{-1}+\tau_{\mathrm{w}}^{-1}\thicksim \tau_{\mathrm{tr}}^{-1}$ means that $\tau_{\mathrm{z}}^{-1}$ is small, since it is expected that $\tau_{\mathrm{w}}^{-1}=\tau_{\mathrm{tr}}^{-1}$. Finally, the fact that $L_{\mathrm{i}}\gg L_{\ast}\thicksim l_{\textrm{tr}}$, together with $\tau_{\mathrm{z}}^{-1}\ll \tau_{\mathrm{tr}}^{-1}$ demonstrate that short-range scattering is not dominant. Moreover, $L_{\Phi\textrm{WL}}$ appears quasi constant between $T=0.35~\mathrm{K}$ and $T=1.5~\mathrm{K}$ around a value that is far below the typical size of the sample. Such saturation of $L_{\Phi\textrm{WL}}$ at low temperature, well below the particle-particle interaction length ($L_{\mathrm{hh}}=\sqrt{D[\frac{\sigma h^{2}}{2\pi e^{2} ln[\sigma h/(2e^{2})]}\frac{1}{k_{\textrm{B}}T}]} =1.6~\mathrm{\mu m}$ at $T=0.35~\mathrm{K}$)\cite{Altshuler1985}, has already been observed in graphene samples\cite{Gorbachev2007} near the CNP. It could be a feature of transport by percolation through electron/hole puddles\cite{Cheianov2007} persisting at finite density (typically $n_{\mathrm{s}}= -2 \times 10^{12}~\mathrm{cm}^{-2} <n^{\ast}$) in the very inhomogeneous BL sample. These results confirm the conclusion drawn from the analysis of the conductivity curves $\sigma(C_{\mathrm{G}}V_{\mathrm{G}}/e)$ that long-range Coulomb scattering by charged impurities trapped in the silicon substrate or in the PMMA top-layer of the graphene-based sample is dominant.

Conductance fluctuations were measured by varying the magnetic induction over a $\pm 1~\mathrm{T}$ magnetic field range with a measurement current of $50~\mathrm{nA}$ at $T=0.35~\mathrm{K}$. The standard deviation is found to be $\delta G_{B}=0.021 e^{2}/h$. In graphene, CF resulting from interference of phase-coherent chiral carrier diffusive paths  are also expected to depend on elastic scattering. In the BL sample, since $L_{\Phi\textrm{WL}} \thicksim L_{\mathrm{i}}$, one expects the amplitude of CF to be properly described by the theory of well-known universal conductance fluctuations (UCF) for diffusive metals\cite{Kharitonov2008,Kechedzhi2008}. Precisely, in the case of a two-dimensional conductor, at a magnetic field larger than the typical magnetic field of WL magnetoresistance, it is given\cite{Lee1985} by $\delta G=0.862\frac{1}{\sqrt{2}}\sqrt{\frac{W}{L}}\frac{min(L_{\Phi},L_{T})}{L}\frac{e^{2}}{h}$ where $L$ and $W$ are the length and the width of the conductor measured, $L_{T}=\sqrt{\hbar D/k_{\textrm{B}}T}$ is the thermal length. Assuming the values $W\approx 4.5~\mathrm{\mu m}$, $L\approx 9~\mathrm{\mu m}$, $T_{\mathrm{eff}}=eRI/k_{\mathrm{B}}=0.87~\mathrm{K}$, one finds $L_{T_{\mathrm{eff}}}=0.36~\mu m < L_{\Phi\textrm{WL}}=0.47~\mu m$ which results in $\delta G=0.018 e^{2}/h$. The good agreement of the experimental magnitude of CF with the theoretical value of the UCF in diffusive metals confirms that $L_{\mathrm{i}}\sim L_{\Phi}\simeq 0.5~\mathrm{\mu m}$, and since $l_{\textrm{tr}}=34~\textrm{nm}\ll L_{\mathrm{i}}$, that long-range scattering is dominant. On the other hand, it shows that conductance fluctuations as a function of the magnetic field are not sensitive to the observed carrier density inhomogeneity or presence of electron and hole puddles.

The analysis of transport at low-magnetic field shows that the dominant mechanism of scattering in our samples is Coulomb interaction with a large concentration of charged impurities closely surrounding graphene flakes (in the silicon substrate and in the PMMA top layer covering the devices). Beyond to drastically reducing the carrier mobility, they are responsible for strong spatial fluctuations of the carrier density that might stay bipolar even at finite density (a few $10^{12}~\mathrm{cm}^{-2}$).

\section{\label{sec:level2}Hall resistance quantization tests of the quantum Hall effect regime}

 \begin{figure}[h]
\begin{center}
\includegraphics[width=8.5cm]{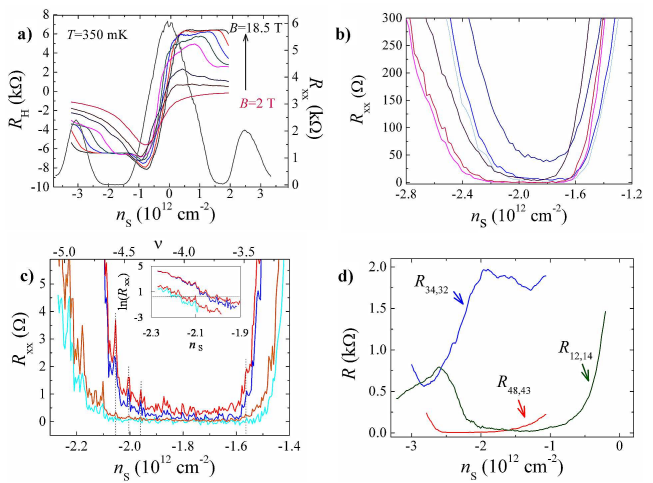}
\caption{a) Hall resistance ($R_{38,24}$) (at magnetic inductions $B=2,~5,~7,~9,~13,~18.5~\mathrm{T}$) and longitudinal resistance ($R_{18,24}$)(at $B=18.5~\mathrm{T}$) in the BL sample at $T=0.35~\textrm{K}$. b) $R_{\mathrm{xx}}=R_{18,24}\times \frac{W}{d_{24}}$ (blue), $R_{\mathrm{xx}}=R_{28,34}\times \frac{W}{d_{34}}$ (magenta) for currents $I=0.5,~1,~3~\mu\mathrm{A}$. c) $R_{\mathrm{xx}}=R_{18,24}\times \frac{W}{d_{24}}$ as a function of $n_{\mathrm{s}}$ around $\nu=-4$ at $T=0.35~\textrm{K}$ and $I=0.5~\mathrm{\mu A}$ (blue); $T=0.35~\mathrm{K}$ and $I=1~\mathrm{\mu A}$ (deep blue);  $T=1.5~\mathrm{K}$ and $I=0.5~\mathrm{\mu A}$ (orange);  $T=1.5~\mathrm{K}$ and $I=1~\mathrm{\mu A}$ (red). Vertical dashed lines underline reproducible fluctuations. Inset: $\mathrm{ln}(R_{\mathrm{xx}})$ as a function of the carrier density. d) Three-terminal resistance of contacts as a function of $n_{\mathrm{s}}$.}\label{fig3}
\end{center}
\end{figure}

In the QHE regime, all measurements were performed using direct current (DC) measurements techniques. Each resistance value reported in the following is the average of values measured for both current directions. About notations, \rx~ is a longitudinal resistance value normalized to a square, for example $R_{\mathrm{xx}}=R_{\mathrm{ij,kl}}\times \frac{W}{d_{\mathrm{kl}}}$ if the longitudinal resistance is measured between terminals $\mathrm{k}$ and $\mathrm{l}$. Fig.~\ref{fig3}a shows Hall and longitudinal resistance as a function of carrier density for the BL sample. Measurements clearly reveal $\nu=\pm 4$ Hall plateaus, typical of the QHE in bilayer graphene, becoming well defined at the highest magnetic inductions. $\nu=\pm 8$ plateaus are barely visible. We note that the energy gap between the lowest LLs ($n=0,1$ and $n=2$) is $92~\mathrm{meV}$ ($1068~\mathrm{K}$ equivalent temperature) at $B=18.5~\mathrm{T}$, thus $3050$ times the thermal energy at $T=0.35~\textrm{K}$. The longitudinal resistance \rx~reported at $B=18.5~\mathrm{T}$ exhibits a central peak corresponding to the degenerate $n=0$ and $n=1$ LLs and minima occuring simultaneously with Hall plateaus. We only investigated the physics of the $\nu=-4$ plateau for holes, which is more flat and characterized by a drop to zero of \rx. Dissipation level in the 2DEG and quality of contacts are essential quantization criteria of the QHE, as demonstrated by several experimental works\cite{Delahaye2003} as well as the Landauer-B\"uttiker theory\cite{Buttiker1988}. The quantization is indeed directly related to the absence of dissipation (\textit{i.e.} of backscattering), the rate of which can be determined by the measurement of \rx. Fig.~\ref{fig3}b shows the behavior of \rx~with hole density on the $\nu=-4$ plateau in the BL sample for several current values increasing from $0.5~\mathrm{\mu A}$ to $5~\mathrm{\mu A}$. The \rx~plateau shrinks and simultaneously the \rx~minimum increases. Fig.~\ref{fig3}b also shows that position and magnitude of \rx~minima depend on the sample region measured. Position variation can be attributed to carrier density fluctuations with a magnitude of a few $10^{11}~\mathrm{cm^{-2}}$ caused by charged impurities, as already mentioned in Section II. In addition, the magnitude variation illustrates that the ignition of QHE breakdown is a very spatially inhomogeneous phenomenon. Fig.~\ref{fig3}c shows that the temperature effect on \rx~ between $0.35~\textrm{K}$ and $1.5~\textrm{K}$ is smaller than the current effect between $0.5~\mu\mathrm{A}$ and $1~\mu\mathrm{A}$. It also shows that \rx~has reproducible fluctuations as a function of $n_{\mathrm{s}}$ with a similar pattern at the two different temperatures and currents. We will later discuss the origin of these fluctuations, particularly visible a bit away from the minimum because of a better signal to noise ratio. Averaging fluctuations (and noise) of \rx~ around specific density values gives typical and relevant mean values of the longitudinal resistance $\bar{R}_{\mathrm{xx}}$. At $T=0.35~\mathrm{K}$ and $I=0.5~\mu\mathrm{A}$, $\bar{R}_{\mathrm{xx}}=\bar{R}_{18,24}\times \frac{W}{d_{24}}$ is $(2 \pm 14) ~\mathrm{m}\Omega$ and $(62 \pm 9)~\mathrm{m}\Omega$ at $n_{\mathrm{s}}=-1.88\times 10^{12}~\mathrm{cm^{-2}}$ and $n_{\mathrm{s}}=-2.01\times 10^{12}~\mathrm{cm^{-2}}$ respectively. These resistance values are to be compared with $100~\mathrm{\mu\Omega}$, the typical value of \rx~which ensures a $10^{-9}$ \rh~accuracy in usual GaAs-based quantum standards (LEP514~\cite{Piquemal1991}).
Contact quality was determined by performing three-terminal measurements of resistance. $R_{\mathrm{3T}}$ (\textit{e. g.} $R_{\mathrm{ij,il}}$) gives the resistance value of the contact $R_{\mathrm{c}}$ (\textit{e. g.} $\mathrm{i}$) combined with a \rx~(\textit{e. g.} $R_{\mathrm{kj,il}}$) contribution. For a good contact, the drop to a negligible value of \rx~ ($\ll 1~\Omega$) in the dissipation-less state leads to a flat minimum of the resistance $R_{\mathrm{3T}}$ giving an upper bound of $R_{\mathrm{c}}$. As observed in Fig.~\ref{fig3}d, for the good Ti/Au contacts of the BL sample, $R_\mathrm{c}$ values deduced from $R_{\mathrm{3T}}$ minima can be as low as $10~\Omega$ (see Fig.~\ref{fig1}a). Note that $R_{\mathrm{3T}}$ minima occur at slightly different $n_{\mathrm{s}}$ values due to the carrier density spatial inhomogeneity. $R_{\mathrm{3T}}$ for contact 3 does not exhibit such a flat minimum with a value higher than $428~\Omega$. The highest resistance value was found equal to $5.9~k\Omega$ for contact 8. These anomalous behaviors can be explained by a large fluctuation of $n_{\mathrm{s}}$ in the voltage arm thin channel ($2~\mu m$) or even by a partial breaking of the constriction probably caused by the sample cooling down too fast. The complete breaking can account for the infinite resistance observed for some other contacts. Contacts 3 and 8 were used as current contacts, rather than voltage, for the Hall resistance precision measurements. It was indeed demonstrated\cite{Jeckelmann1997} that a very resistive detecting voltage contact can lead to a deviation from quantization notably because being unable to restore the equilibrium of the edge state population\cite{Buttiker1988}. Although we used Pd instead of Ti/Au to make contact to graphene, similar observations are reported in the ML sample. The five contacts used to perform measurements have low resistance values ranging from $15~\Omega$ to $260~\Omega$ (see Fig.~\ref{fig1}b).

\begin{figure}[h]
\begin{center}
\includegraphics[width=8.5cm]{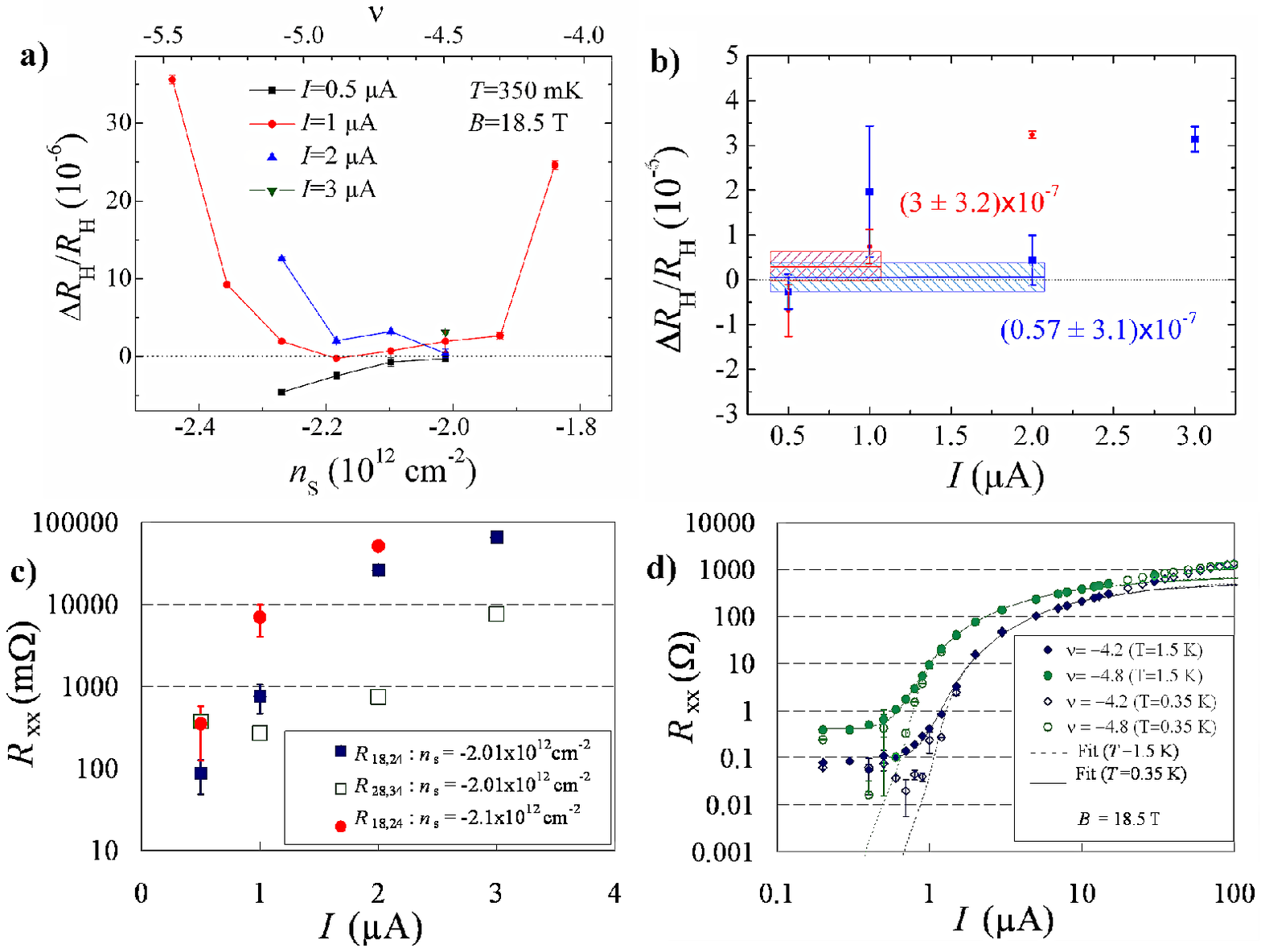}
\caption{a) Relative Hall discrepancy $\Delta$\rh/\rh~as afunction of $n_{\mathrm{s}}$ (and $\nu$ in upper-scale) at four currents. b) $\Delta$\rh/\rh~as a function of $I$ at $n_{\mathrm{s}} = -2.01\times 10^{12}~\textrm{cm}^{-2}$ (blue) and at $n_{\mathrm{s}} = -2.1\times 10^{12}~\textrm{cm}^{-2}$ (red). c) $R_{\mathrm{xx}}=R_{18,24}\times \frac{W}{d_{24}}$ as a function of $I$ at $n_{\mathrm{s}} = -2.01\times 10^{12}~\textrm{cm}^{-2}$ (filled blue square) and at $n_{\mathrm{s}} = -2.1\times 10^{12}~\textrm{cm}^{-2}$ (filled red circle), $R_{\mathrm{xx}}=R_{28,34}\times \frac{W}{d_{34}}$ as a function of $I$ at $n_{\mathrm{s}} = -2.01\times 10^{12}~\textrm{cm}^{-2}$ (unfilled green square). d) $R_{\mathrm{xx}}=R_{18,24}\times \frac{W}{d_{24}}$ as a function of $I$ for filling factors $\nu = -4.2$ (blue diamond) and $\nu = -4.8$ (green circle) at temperatures $T=1.5~\mathrm{K}$ (filled symbols) and $T=0.35~\mathrm{K}$ (unfilled symbols). Error bars correspond to uncertainties given within one standard deviation, $1\sigma$.}\label{fig4}
\end{center}
\end{figure}

\begin{figure}[h]
\begin{center}
\includegraphics[width=8.5cm]{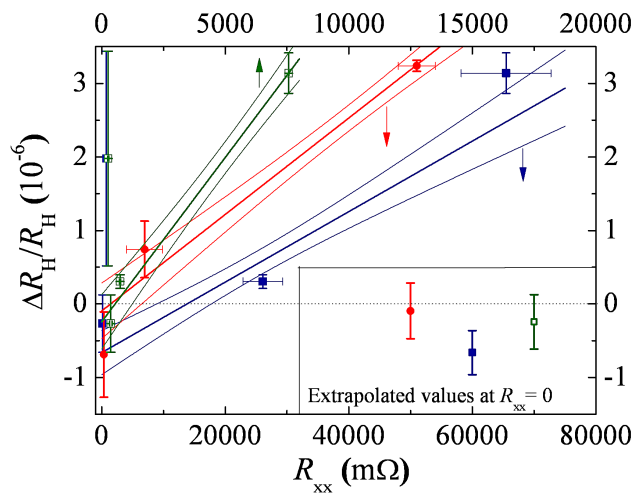}
\caption{$\Delta$\rh/\rh~as a function of $R_{\mathrm{xx}}=R_{18,24}\times \frac{W}{d_{24}}$ at $n_{\mathrm{s}}=-2.01\times 10^{12}~\mathrm{cm^{-2}}$ (filled blue square), $\Delta$\rh/\rh~as a function of $R_{\mathrm{xx}}=R_{28,34}\times \frac{W}{d_{34}}$ at $n_{\mathrm{s}}=-2.01\times 10^{12}~\mathrm{cm^{-2}}$ (unfilled green square), $\Delta$\rh/\rh~as a function of $R_{\mathrm{xx}}=R_{18,24}\times \frac{W}{d_{24}}$ at $n_{\mathrm{s}}=-2.1\times 10^{12}~\mathrm{cm^{-2}}$ (filled red circle). Errors bars correspond to measurement uncertainties given within one standard deviation, $1\sigma$.}\label{fig5}
\end{center}
\end{figure}

We then performed accurate measurements of \rh~in terms of \rk~using a resistance bridge equipped with a SQUID based cryogenic current comparator. In practice, the Hall resistance is compared to a well-known $100~\Omega$ wire resistor calibrated in terms of a GaAs based quantum resistance standard (LEP514).
In the BL sample, Fig.~\ref{fig4}a reports the relative deviation of \rh~=~$R_{38,24}$ from its nominal value $\Delta$\rh/\rh~=~\rh/(\rk/4)-1 as a function of $n_{\mathrm{s}}$. All uncertainties are given within one standard deviation ($1\sigma$). Let us note that the resistance measured not only includes a pure transverse resistance but also a longitudinal resistance contribution, because the line between voltage terminals is not perpendicular to the one between current terminals. Measurements clearly show a flat resistance plateau within 3 parts in $10^{6}$ over a $2\times 10^{11}~\mathrm{cm^{-2}}$ carrier density range when measured with a current below $1~\mu\mathrm{A}$. At the lowest measurement current $I=0.5~\mu\mathrm{A}$, deviations from quantization at highest carrier density agree with the expected shape of the Hall plateau (decrease of resistance on plateau edges). The shape evolution at higher currents is attributed to a \rx~contribution which adds to the transverse resistance and increases with the current. This coupling between \rh~and \rx, which always exists to some extend in GaAs based quantum resistance standard\cite{Cage1984}, will be later discussed in more details. The flatness appears worse at $I=2~\mu\mathrm{A}$, as expected with regards to the large increase of \rx, fluctuating with carrier density as previously discussed. Fig.~\ref{fig4}b confirms that deviations from quantization start to drastically increase from $I=2~\mu\mathrm{A}$ at $n_{\mathrm{s}}=-2.1\times 10^{12}~\mathrm{cm^{-2}}$ and from $I=3~\mu\mathrm{A}$ at $n_{\mathrm {s}}=-2.01\times 10^{12}~\mathrm{cm^{-2}}$. As demonstrated in Fig.~\ref{fig4}c, the increase of deviation due to current is accompanied by a large increase of \rx~at both densities. The weighted mean value of $\Delta$\rh/\rh~values measured at currents below these critical currents leads to small deviations of $(0.57\pm 3.1)\times 10^{-7}$ and $(3.0\pm 3.2)\times 10^{-7}$ at $n_{\mathrm {s}}=-2.01\times 10^{12}~\mathrm{cm^{-2}}$ and $n_{\mathrm {s}}=-2.1\times 10^{12}~\mathrm{cm^{-2}}$ respectively. Since \rx~is the relevant parameter of quantization, $\Delta$\rh/\rh~as a function of \rx~is then reported in Fig.~\ref{fig5} from data of \ref{fig4}b and \ref{fig4}c for the two carrier densities. Although dissipation is inhomogeneous in the sample, as the very different values of \rx~measured using voltage terminal-pairs (2,4) and (3,4) at $n_{\mathrm {s}}=-2.01\times 10^{11}~\mathrm{cm^{-2}}$ express again, all deviations scale quite linearly with \rx, indicating a common coupling mechanism between Hall and longitudinal resistances. This linear relationship, which is usually observed in GaAs based quantum resistance standards, is generally explained in terms of an effective misalignment of Hall probes, either due to a lack of carrier density homogeneity\cite{VanderwelPhd1988,Vanderwel1988} in the sample or to current flow chiral nature in finite width voltage terminals\cite{DomingezPhd1988}. In a good quantum Hall resistance standard, one usually finds $\Delta$\rh/\rh=$\alpha$\rx/\rh~with $\alpha \simeq 0.1-1$. In our case voltage terminals are really misaligned, which should lead to a unity coupling factor. But from slopes we deduce $\alpha$ values in the range $10^{-2}$ to $10^{-4}$, depending where \rx~is measured. This means that, due to inhomogeneity, \rx~values are not quantitative measurements of the dissipation level between Hall probes 2 and 4 when the current flows between terminals 3 and 8. Nevertheless, the values as a whole give a qualitative representation of the dissipation current behavior in the sample. It is therefore justified to extrapolate $\Delta$\rh/\rh~in the dissipationless limit (\rx$=0$) at which the perfect quantization is expected. In this limit, at $n_{\mathrm {s}}=-2.01\times 10^{12}~\mathrm{cm^{-2}}$, we find $\Delta R_{\mathrm{H}}/R_{\mathrm{H}}(R_{\mathrm{xx}}=0)=(-6.62\pm 3.0)\times 10^{-7}$ and $\Delta R_{\mathrm{H}}/R_{\mathrm{H}}(R_{\mathrm{xx}}=0)=(-2.43\pm 3.7)\times 10^{-7}$ using \rx~ measurement with voltage terminal pairs (2,4) and (3,4) respectively. Agreement of these two values within the measurement uncertainty corroborates our extrapolation protocol. At $n_{\mathrm {s}}=-2.1\times 10^{12}~\mathrm{cm^{-2}}$, $\Delta R_{\mathrm{H}}/R_{\mathrm{H}}(R_{\mathrm{xx}}=0)=(-0.94~\pm~3.78)\times 10^{-7}$, thus the Hall resistance stays quantized within the measurement uncertainty. But the carrier density value $-2.01~\times~10^{12}~\mathrm{cm^{-2}}$ seems to ensure a minimal sensitivity of the Hall resistance to dissipation.

\begin{figure}[h]
\begin{center}
\includegraphics[width=8.5cm]{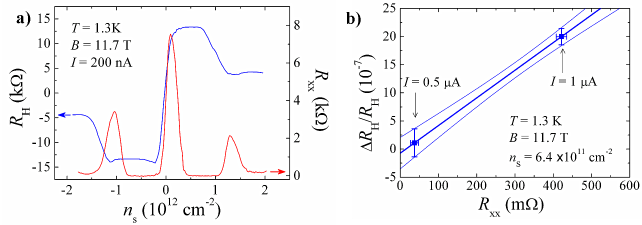}
\caption{a) Hall (\rh~$=R_{06,39}$) and longitudinal ($R_{\mathrm{xx}}=R_{06,23}\times \frac{W}{d_{23}}$) resistances as a function of $n_{\mathrm{s}}$ at $B=11.7~\mathrm{T}$, $I=200~\mathrm{nA}$ and $T=1.3~\mathrm{K}$.  b) $\Delta$\rh/\rh~ as a function of~\rx~at $n_{\mathrm{s}}=6.4\times 10^{11}~\mathrm{cm^{-2}}$. Error bars correspond to uncertainties given within one standard deviation, $1\sigma$.}\label{fig6}
\end{center}
\end{figure}

A similar study was carried out on the ML sample. Fig.~\ref{fig6}a shows $\nu=\pm2$ and $\nu=\pm6$ Hall plateaus at $B=11.7~\mathrm{T}$ and $T=1.3~\mathrm{K}$ that are typical of half-integer QHE in monolayer graphene. Fig.~\ref{fig6}b shows two couples of $\Delta$\rh/\rh~=~\rh/(\rk/2)-1 and \rx~values measured with two measurement currents $0.5~\mu\mathrm{A}$ and $1~\mu\mathrm{A}$ at $n_{\mathrm {s}}=6.4\times 10^{11}~\mathrm{cm^{-2}}$ on the $\nu=2$ plateau. Although the deviation strongly increases from $I=1~\mu\mathrm{A}$, the extrapolation to zero dissipation gives $\Delta R_{\mathrm{H}}/R_{\mathrm{H}}(R_{\mathrm{xx}}=0)=(0.73 \pm 2.8)\times 10^{-7}$. The degree of accuracy achieved in the ML and BL samples is therefore similar. It is independent of the ratio of the energy gap to thermal energy since $\frac{92~\mathrm{meV}}{0.35~\mathrm{K}\times k_{\textrm{B}}}$ in the BL sample is by $2.8$ higher than $\frac{123~\mathrm{meV}}{1.3~\mathrm{K}\times k_{\textrm{B}}}$ in the ML sample. The quantization accuracy is probably determined by the presence of the same high concentration of charged impurities in both samples leading to the carrier density inhomogeneity and low carrier mobility. The impact of charged impurities on the QHE breakdown will be discussed in the following through detailed analysis of the current dependence of \rx~in the BL sample.

\section{\label{sec:level2} Dissipation mechanism in the QHE regime in bilayer graphene}

\subsection{\label{sec:level2} Current dependence of the longitudinal resistance}

Dissipation in GaAs/AlGaAs 2DEG was found to increase with temperature or current through several mechanisms. At low temperature and low current, carriers can backscatter from one edge to the opposite edge through localized states by variable range hopping (VRH) with soft Coulomb gap, characterized by a temperature behavior of the conductivity $(\sigma_{0\mathrm{VRH}}/T)\exp[-(T_{0}(\xi_{\mathrm{loc}})/T)^{1/2}]$ where $k_{\mathrm{B}}T_{0}(\xi_{\mathrm{loc}})= e^2/(4\pi\varepsilon_0\varepsilon_r\xi_{\mathrm{loc}})$ and $\xi_{\mathrm{loc}}$ is the localization length\cite{Shklovskii1984,Polyakovprl1993,Furlan1998} a lower bound of which is the magnetic length $l_B=\sqrt{\hbar/eB}$. Current effect manifests itself as an effective temperature $k_{\mathrm{B}}T_{\mathrm{eff}}=eV_{\mathrm{H}}\xi_{\mathrm{loc}}/W$. At a higher temperature, conductivity is activated following the behavior $\sigma_0\exp[-(T_{\mathrm{Act}}/T)]$, where $\sigma_0$ is close to $e^2/h$ and weakly dependent on the electron-phonon coupling in case of a short-range potential\cite{Polyakovprl1994} but expected to be universal and equal to $2e^2/h$ in case of a long-range potential\cite{Polyakovprl1994U}. $T_\mathrm{Act}$ is typically related to the cyclotron gap. Experimentally, VRH mechanism was also observed in monolayer graphene\cite{Giesbergvrh2009,Glattli,Tzalenchuk2011}. In samples based on exfoliated graphene transferred on $\textrm{Si/SiO}_{2}$ substrate, screening of the Coulomb interaction by the close metallic back-gate even restores the usual two-dimensional VRH mechanism with a temperature dependence $\exp[-(T_{1}(\xi_{\mathrm{loc}})/T)^{1/3}]$ where $T_{1}(\xi_{\mathrm{loc}})\sim 1/(g(E_\textrm{F}){\xi_{\mathrm{loc}}^2})$ and $g(E_\textrm{F})$ is the density of states\cite{Glattli}. Conductivity activation by temperature was also observed in graphene systems\cite{Giesbergact2007, Giesbergact2009}.

On the other hand, there are few reports\cite{Singh2009, Glattli} dealing with detailed investigation of the QHE breakdown by increasing the current in exfoliated graphene. For semiconductors 2DEGs, several electric field assisted mechanisms have been considered to explain the large increase of longitudinal conductivity leading to the QHE breakdown\cite{Cage1983}:  quasi-elastic inter-Landau levels scattering (QUILLS)\cite{Heinonen1984,Eaves1986} possibly combined with intra Landau levels scattering\cite{Chaubet1995, Chaubet1998}, increase of delocalized electron states in Landau levels\cite{Trugman1983}, ordinary\cite{Ebert1983} electron heating, bootstrap-type\cite{Komiyama1985, Komiyama1996} electron heating (particularly efficient in large-size samples), and electron percolation between sample edges by merging of compressible islands\cite{Tsemekhman1997}. In a sample made of exfoliated graphene on $\textrm{Si/SiO}_{2}$ substrate, Singh and co-workers\cite{Singh2009} deduced from the measurement of breakdown current dependence on integer filling factor that the QHE regime is broken by inter-Landau levels scattering in presence of large local electric field.

Fig.~\ref{fig4}d  reports on \rx~dependence on current measured at two filling factors $\nu$ (or $n_{\mathrm{s}}$ values) near $\nu=-4$ in the BL sample under a magnetic induction of $18.5~\textrm{T}$ and for both temperatures $0.35~\textrm{K}$ and $1.5~\textrm{K}$. It displays exponential increases of \rx~ over three orders of magnitude above a critical current.
More precisely, one can define a breakdown current $I_{\mathrm{c}}$ by the value above which conductivity exceeds $2.10^{-8}~\textrm{S}$. $I_{\mathrm{c}}$ linearly decreases for decreasing $\nu$ values departing from the filling factor $\nu=-4$ in the range from approximately $1.5~\mu\mathrm{A}$ to $0.5~\mu\mathrm{A}$ (see Fig.~\ref{fig8}a). The breakdown current at $T=0.35~\textrm{K}$ is slightly higher than at $T=1.5~\textrm{K}$. This behavior is also observed in GaAs samples\cite{JeckelmannIEEE2001}.

\begin{figure}[b]
\begin{center}
\includegraphics[width=8.5cm]{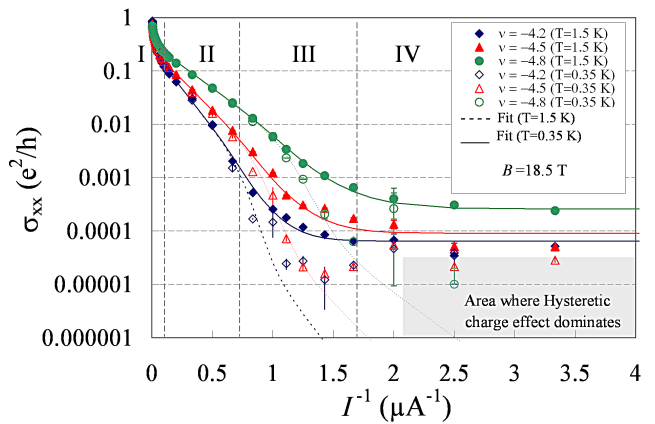}
\caption{Conductivity $\sigma_{\mathrm{xx}}$ ($e^{2}/h$) as a function of $I^{-1}$ at $\nu=-4.2$ (blue diamond), $\nu=-4.5$ (red triangle), $\nu=-4.8$ (green circle) (filled at $T=1.5~\textrm{K}$ and unfilled at $T=0.35 ~\textrm{K}$). Solid lines and dots lines correspond to theoretical adjustments at $T=1.5~\textrm{K}$ and $T=0.35~\textrm{K}$ respectively. Error bars correspond to uncertainties given within one standard deviation, $1\sigma$.}\label{fig7}
\end{center}
\end{figure}

Fig.~\ref{fig7} clearly displays the existence of four current regimes for the conductivity calculated by $\sigma_{\mathrm{xx}}= R_{\mathrm{xx}}/(R_{\mathrm{xx}}^{2}+R_{\mathrm{H}}^{2})$. We will later discuss the first regime I for very high currents. For currents down to $1.5~\mu\mathrm{A}$ (second current regime II), conductivity $\sigma_{\mathrm{xx}}$ decreases by decreasing the current following a unique phenomenological fitting function $\sigma_{0,\nu}\exp[-\Delta E_{\mathrm{a}}(\nu)/eR_{\mathrm{H}}I]$ at both temperatures. $\sigma_{0,\nu}$ is found quite universal around $0.25 {e^2}/h$ within $30\%$ for all $\nu$ values. Fig.~\ref{fig8}a shows that $\Delta E_{\mathrm{a}}(\nu)$ scales linearly with $\nu$, similarly to $\Delta E_{\mathrm{th}}(\nu)= (\sqrt{2}\hbar\omega_{\mathrm{c}}/2)(1+(\nu+4)/2)$ which is the energy difference between the Fermi level and the center of the $n=-2$ Landau level if a constant density of states is assumed. The small discrepancy between $\Delta E_{\mathrm{th}}(\nu)$ and $\Delta E_{\mathrm{a}} (\nu)$ results in a deviation of the $\nu$ value for which $\Delta E_{\mathrm{a}}=0$ from $-6$, center of the $n=-2$ LL. The resulting filling factor $\nu_{\mathrm{edge}}$ can be interpreted as the mobility edge which separates localized and extended states near the center of the $n=-2$ LL. At lower current in the third current regime III, $\sigma_{\mathrm{xx}}$ at $T=0.35~\mathrm{K}$ decreases more quickly with $I$ decreasing and departs from $\sigma_{\mathrm{xx}}$ at $T=1.5~\mathrm{K}$ that roughly continues to follow the law characterizing regime II. Reducing the current below $0.7~\mathrm{\mu A}$ leads to the fourth regime IV where conductivity apparently saturates at values $\sigma_{T,\nu}$ different for the two temperatures. At $T=0.35~\mathrm{K}$, the conductivity threshold cannot be determined because of the increasing weight of some hysteretic charging effect altering measurements of $\sigma_{\mathrm{xx}}$ for currents below $0.5~\mathrm{\mu A}$. Consequently, the reasonable assumption that $\sigma_{T,\nu}$ follows the typical temperature dependence of VRH mechanism cannot be confirmed.

We remark that the monotonous behavior of conductivity following the current dependence $\sigma_{0,\nu}\exp[-\Delta E_{\mathrm{a}}(\nu)/eR_{H}I]$ dominates at $T=1.5~\textrm{K}$ in regime II and III over more than three conductivity decades down to $I=0.7~\mu\mathrm{A}$ (and at $T=0.35~\textrm{K}$ in regime II), and cannot be explained by an activation effect caused by a simple heating of electrons by current since it does not manifest itself at the lowest temperature $T=0.35~\mathrm{K}$ in the low current regime III down to the same value $I=0.7~\mu\mathrm{A}$. More quantitatively, a conductivity increase due to heating by current should be described by $\sigma_{0,\nu}\exp[-\Delta E_{\mathrm{th}}(\nu)/k_\mathrm{B} T_{\mathrm{el}}]$ with $T_{\mathrm{el}}$ the effective electron temperature resulting from the heating. The correct adjustment of data at $T=1.5~\textrm{K}$ approximately above $I=0.7~\mu\mathrm{A}$ by $\sigma_{0,\nu}\exp[-\Delta E_{\mathrm{th}}(\nu)/eR_{\mathrm{H}}I]$ would result in $k_\mathrm{B} T_{\mathrm{el}}=eV_{\mathrm{H}}=eR_{\mathrm{H}}I$, leading to an effective temperature $T_{\mathrm{el}}$ of $52~\mathrm{K}$ for $I=0.7~\mu\mathrm{A}$. This is not in agreement with the electronic temperature, which is obviously close to $1.5~\mathrm{K}$ since conductivity starts to be current independent below $I=0.7~\mu\mathrm{A}$, with a constant value expected to be determined by the bath temperature. Finally, absence of strong asymmetry of \rx~values (there are similar within $30\%$) with respect to current direction indicates that there is no strong local electron heating in current contact\cite{Meziani2004}. This rules out any strong role of current contacts 8 ($5.9~\mathrm{k\Omega}$) and 3 ($< 5.9~\mathrm{k\Omega}$) on the observed breakdown mechanism.

The exponential dependence of conductivity on $\Delta E_{\mathrm{th}}(\nu)$ in current regime II rather directs towards a dissipation mechanism based on quasi-elastic inter-Landau levels scattering (QUILLS) assisted by the electric field. In current regime III, the decrease of conductivity at $T=0.35~\mathrm{K}$ suggests that the QUILLS mechanism is combined with a blockade mechanism manifesting itself approximately below $T=1.5~\mathrm{K}$ and a threshold current ($\sim 1~\mu\mathrm{A}$), like a heating mechanism by current. It appears that all conductivity curves under regimes II, III and IV, at both $T=1.5~\mathrm{K}$ and $T=0.35~\mathrm{K}$, can be adjusted (see Fig.~\ref{fig7} and Fig.~\ref{fig4}d) by a unique fitting function $\sigma_{\mathrm{xx}}=\sigma_{T,\nu}+\sigma_{0,\nu}\exp[-\Delta E_{\mathrm{a}}(\nu)/eR_{\mathrm{H}}I]\exp[-E_{\mathrm{c}}/(k_\mathrm{B}(T+\gamma \sigma_{\mathrm{xx}}{V_\mathrm{H}}^2))]$. $\sigma_{0,\nu}\simeq 0.25 {e^2}/h$ for all $\nu$ values, $E_{\mathrm{c}}=95~\mathrm{\mu eV}$, $\gamma=0.48~\textrm{K/pW}$ and $\Delta E_{\mathrm{a}}$ near $\Delta E_{\mathrm{th}}$ as already explained. The contribution $\sigma_{T,\nu}$ is chosen to adjust conductivity in the low current regime IV only at $T=1.5~\mathrm{K}$.

\begin{figure}[h]
\begin{center}
\includegraphics[width=8.5cm]{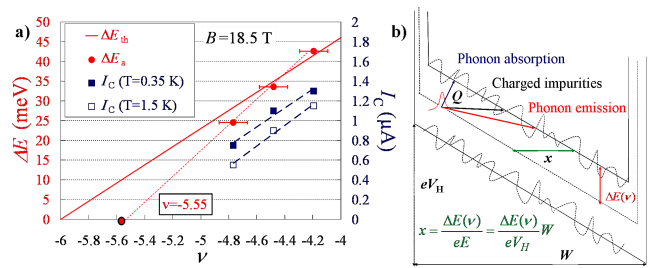}
\caption{a) $\Delta E_{\mathrm{th}}$ ($\mathrm{meV}$) as a function of $\nu$ (red solid line), $\Delta E_\mathrm{a}$ ($\mathrm{meV}$) as a function of $\nu$ (red filled circle), breakdown currents $I_{\mathrm{c}}$ as a function of $\nu$ at $T=0.35~\mathrm{K}$ (blue filled square) and $T=1.5~\mathrm{K}$ (blue unfilled square). Error bars correspond to uncertainties given within one standard deviation, $1\sigma$ b) Schematics of inter-Landau levels transitions in case of disordered 2DEG.}\label{fig8}
\end{center}
\end{figure}

\subsection{\label{sec:level2} Phenomenological model based on Quasi-elastic Inter-Landau Level Scattering (QUILLS)}

In order to explain with the QUILLS mechanism the main current dependence of conductivity observed in regime II, let us first consider an homogenous electric field and harmonic oscillator wave functions for the carriers. The tilting of LLs by the electric field brings closer localized states at Fermi energy and extended states in the nearest Landau levels, increases the wavefunction overlap, and thus leads to an increased transition probability $P$ between LLs. Through scattering processes by phonons and/or charged impurities, $P$ is proportional to the wavefunction overlap given by $\exp[-{Q^2}{l_B}^2]$ where $Q$ is the typical direct momentum between Landau levels $Q=\omega_{\mathrm{c}}B/E$ (see fig. 8b). This should lead to $P\propto \exp[-(\Delta E_{\mathrm{th}}(\nu)/eEl_B)^2]=\exp[-(\Delta E_{\mathrm{th}}(\nu)/eV_{\mathrm{H}})^2(W/l_B)^2]$\cite{Chaubet1998}, thus a transition probability different from the one observed. But, in presence of disorder, at a length scale larger than $l_B$, one expects a dependence $\exp[-x/\xi_{\mathrm{loc}}(\nu)]$ of the localized state wavefunction tail where $\xi_{\mathrm{loc}}(\nu)$ is the localization length varying like $(\nu-\nu_{\mathrm{c}})^{-(2.3\pm 0.1)}$ with $\nu_{\mathrm{c}}$ the filling factor of the Landau level center\cite{Yoshioka1998}. This case should be particularly valid in bilayer graphene where the energy gap between LLs is very large. From the distance $x=\Delta E_{\mathrm{th}}(\nu)/eE$ between an initial localized state at the Fermi energy and a final extended state in the Landau level at the same energy, one therefore expects $P\propto \exp[-\Delta E_{\mathrm{th}}(\nu)/(eE\xi_{\mathrm{loc}})]$. This model can well describe the main exponential current dependence observed in current regime II provided that large local electric fields with a magnitude around $V_{\mathrm{H}}/\xi_{\mathrm{loc}}$ are considered. Assuming $\xi_{\mathrm{loc}} \sim l_B =6~\mathrm{nm}$ at $\nu=-4$ and $B=18.5\mathrm{T}$ actually leads to a high value of the electric field $V_{\mathrm{H}}/\xi_{\mathrm{loc}} \sim 10^{6}~\textrm{V/m}$ for $I=1~\mathrm{\mu A}$. Besides, the correct adjustment of data by this model needs $\Delta E_{\mathrm{th}}$ to be replaced by $\Delta E_{\mathrm{a}}$ that can be interpreted as the energy difference between the Fermi energy and the mobility edge of the nearest LL.

We argue that the high concentration of charged impurities ($2.10^{12}~\mathrm{cm^{-2}}$) in the substrate can lead to such large electric fields. In absence of magnetic field, it was demonstrated in section II that charged impurities create carrier density fluctuations with a magnitude of $10^{12}~\mathrm{cm^{-2}}$ and a typical correlation length $\xi=11~\textrm{nm}$ in the considered BL sample notably manifesting themselves as electron and hole puddles near the CNP. These fluctuations are combined with more macroscopic carrier density variations extending over larger spatial scales. Their impact at high magnetic field in the QHE regime has been addressed. Scanning of graphene on $\textrm{Si/SiO}_{2}$ substrate by tunneling spectroscopy\cite{Jung2011} or single electron transistor technique\cite{Martin2009} has indeed shown that the potential landscape drawn by charged impurities is partially screened by the Coulomb interaction and leads to the existence of compressible islands surrounded by incompressible strips like in AlGaAs/GaAs 2DEG\cite{Ilani2004}. Jung and co-authors\cite{Jung2011} even show that electron or hole puddles at zero magnetic field turn into compressible islands surrounded by incompressible strips in the QHE regime. It turns out that the localization length $\xi_{\mathrm{loc}}$, or rather its lower bound $l_{B}$, as well as the characteristic length of incompressible strips across which Hall potential drops, could be similar to the electron and hole puddle correlation length $\xi$. Thus, the existence of large local electric field in the BL sample with a typical magnitude $V_{\mathrm{H}}/\xi \sim 10^{6}~\textrm{V/m}$ should result from the strong carrier density fluctuations caused by large concentration of charged impurities. Similar explanation was proposed by Sing and co-workers\cite{Singh2009}.
Another way to understand the impact of the carrier density fluctuations is to consider that they turn into spatial variations of the filling factor in the QHE regime. Otherwise, the current flows along a path minimizing the dissipation that is expected to occur at $\nu=-4$. Given the correlation length of the filling factor (or similarly of carrier density) fluctuations and the small width of the sample, it is therefore likely that the current flows along a narrow percolating incompressible path having a typical width $\xi=11~\textrm{nm}$. The potential drop concentration across this path leads to the existence of large local electric fields. Beyond the enhancement of the electric field, the role of charged impurities in the QHE breakdown has been investigated in conventional semiconductor heterostructure. While charged impurities are kept away from the 2DEG by the $10~\textrm{nm}$ to $40~\textrm{nm}$ thick spacer, acoustic electron-phonon interaction controls the QHE breakdown because elastic scattering by ionized impurities increases the inter-Landau level transition rate at higher electric field. But numerical work\cite{Chaubet1995} shows that the closer charged impurities are from the 2DEG the lower the electric field at which they are efficient. We therefore propose that a high concentration of charged impurities located at only about $1~\textrm{nm}$ from graphene in the BL sample could itself be responsible for inter-Landau level transitions, which are in addition enhanced by the strong electric fields introduced by the carrier density inhomogeneity these impurities induce. This results in QHE breakdown currents (typically $0.2~\textrm{A/m}$) that are low as compared to expectations in graphene from large LL energy gap and prevents from observing backscattering by VRH at currents above $I\approx 0.7~\mu\mathrm{A}$. On the other hand, in samples made from exfoliated monolayer graphene of higher mobility where short-range scatterers dominate transport at low magnetic field\cite{Monteverde2010}, dissipation in the QHE regime was observed\cite{Glattli} to occur through VRH when increasing current up to $\approx 30~\mu\mathrm{A}$.

The term $\exp[-E_{\mathrm{c}}/(k_{\mathrm{B}}(T+\gamma \sigma_{\mathrm{xx}}{V_{\mathrm{H}}}^2))]$ allows the description of the temperature effect and the weak heating effect by current, clearly visible in current regime III (below $1~\mathrm{\mu A}$). It phenomenologically models a blockade mechanism that can be activated by thermal energy above a critical energy $E_{\mathrm{c}}=95~\mathrm{\mu eV}$. The effective temperature of carriers given by $T^{\ast}=T+\gamma \sigma_{\mathrm{xx}}{V_{\mathrm{H}}}^2$ leads to the best adjustment of data notably reproducing very well the sharpness of the crossover between large and low current regimes at $T=0.35~\mathrm{K}$. Even at $T=1.5~\mathrm{K}$, this exponential term allows a better adjustment of conductivity. The proportionality of temperature increase with the dissipated electric power means that carriers are very badly coupled with phonons of graphene or substrate. This results in a large temperature dependence on power manifested in the value of the parameter $\gamma \sim 0.48~\mathrm{K/pW}$. At $I=1~\mathrm{\mu A}$, $T^{\ast}$ amounts to about $0.5~\mathrm{K}$, $2~\mathrm{K}$, and $5.5~\mathrm{K}$ for $\nu$ values -4.2, -4.5 and -4.8 respectively. The electronic temperature increases all the more so as the $\nu$ value departs from $\nu =-4$, because of the higher mean conductivity leading to more dissipation.
The origin of this blockade mechanism manifesting itself at low temperature and typically clearly visibly below $T=1.5~\mathrm{K}$ in our experiment, is not understood. However, it is worth mentioning that characteristic energies of phonon absorption $\hbar c_{\mathrm{s}}/l_B$, where $c_{\mathrm{s}}$ is the sound velocity, are $22~\mathrm{meV}$ ($25.6~\mathrm{K}$) and $6.6~\mathrm{meV}$ ($7.7~\mathrm{K}$) for phonons of graphene bilayer ($c_{\mathrm{s}}=2\times 10^{4}~\mathrm{ms^{-1}}$) and of $\textrm{Si0}_{2}$ ($c_{\mathrm{s}}=6\times 10^{3}~\mathrm{ms^{-1}}$) respectively, thus well above $95~\mathrm{\mu eV}$. On the other hand, the energy value $1.8~\mathrm{meV}$ ($2.1~\mathrm{K}$) for phonons of PMMA covering the sample ($c_{\mathrm{s}}=1.6\times 10^{3} ~\mathrm{ms^{-1}}$) could be compatible with our observations. An explanation based on Coulomb blockade effect in compressible islands is more improbable since the low value of $E_{\mathrm{c}}$ would mean oversized islands.

Finally, the observed disappearance of the exponential regime in current regime I (see Fig.~\ref{fig4}d) can naturally be explained by the QUILLS mechanism because of the overlap integral saturation occurring when Landau levels are very tilted. At higher currents, conductivity slowly increases with a polynomial dependence $\sigma_{\mathrm{xx}} \propto I^{\beta}$ with $\beta$ varying from 1/3 to 2/3 for $\nu$ values from -4.8 to -4.2.

\begin{figure}[h]
\begin{center}
\includegraphics[width=8.5cm]{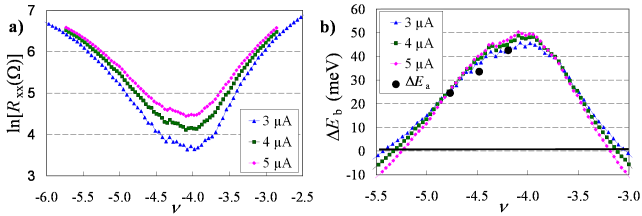}
\caption{a) $\ln R_\mathrm{xx}$ ($R_{18,24}$) as a function of $\nu$ for four current values at $T=0.35~\mathrm{K}$. b) $\Delta E_{\mathrm{b}}=-\ln [R_\mathrm{xx}/(\sigma_{0, \nu} R_\mathrm{H}^{2})]eR_\mathrm{H}I$ as a function of $\nu$; $\Delta E_{\mathrm{a}}$ values are reported as black dot points.}\label{fig9}
\end{center}
\end{figure}

The linear dependence of $\Delta E_{\mathrm{a}}(\nu)$ on $\nu$ demonstrated in the range between $\nu=-4.8$ and $\nu=-4.2$ means that density of states is to be quasi constant. It is possible to verify this hypothesis on a larger range of filling factors $\nu$ from the \rx~dependence on $\nu$ measured at several currents. Fig.~\ref{fig9}a  reports $\ln(R_{\mathrm{xx}})$ as a function of $\nu$ measured with current values $3,~4,~5~\mathrm{\mu A}$ for which the term $\exp[-E_{\mathrm{c}}/(k_{\mathrm{B}}(T+\gamma \sigma_{\mathrm{xx}}{V_{\mathrm{H}}}^2))]\sim 1$ has no impact. Fig.~\ref{fig9}b shows that all four curves displaying $\Delta E_{\mathrm{b}}=-\ln [R_{\mathrm{xx}}/(\sigma_{0, \nu} R_\mathrm{H}^{2})]eR_\mathrm{H}I$ (with $\sigma_{0, \nu}=0.25e^2/h$) approximatively merge into a unique curve, except far from $\nu=-4$. This nicely shows that conductivity well follows the current dependence $\sigma_{0,\nu}\exp[-\Delta E_{\mathrm{b}}(\nu)/eR_{H}I]$ and reinforces the meaning of $\Delta E_{\mathrm{b}}(\nu)$ as the energy difference between the Fermi energy and the mobility edge of the nearest Landau levels. $\Delta E_{\mathrm{b}}(\nu)$ draws the dependence of this energy difference on $\nu$. This energy reaches a maximum value of $45~\mathrm{meV}$ at exactly $\nu=-4$, which is half the energy gap as expected. Fig.~\ref{fig9}b first shows that $\Delta E_\mathrm{a}$ values deduced at $\nu=-4.2,~-4.5,~-4.8$ from the adjustment of the current dependence of conductivity perfectly match the $\Delta E_{\mathrm{b}}(\nu)$ curve deduced from the filling factor dependence of \rx~at different currents. Second, it shows that $\Delta E_{\mathrm{b}}(\nu)$ linearity holds on both side of $\nu=-4$ over more than one unit variation. A constant density of states, as deduced, could be a consequence of the large Landau level overlap inherent to low carrier mobility. Far from $\nu=-4$, curves do not superimpose in a unique curve which means that QUILLS is no more the mechanism responsible for conductivity. The sub-linearity of $\ln(R_{\mathrm{xx}})$ as observed in Fig.~\ref{fig9}a can rather be explained by a saturation of the wavefunction overlap at filling factors near mobility edges. Therefore, even if we expect an increase of density of states, the energy determined near mobility edges in Fig.~\ref{fig9}b is not relevant. Extrapolating the linear behavior of $\Delta E_{\mathrm{b}}(\nu)$ at zero energy should give a reasonable estimate of the mobility edge filling factor of the $n=-2$ Landau level $\nu_{\mathrm{edge}(n=-2)}=-5.55$. This value means that the mobility edge energy should depart from the ($n=-2$) Landau level energy by $10.35 \pm 2.3~\textrm{meV}$ ($120~\textrm{K}$). This value can be compared with the half width of Landau level predicted by the Born approximation to be equal to $\hbar / 2\tau_{e}$. It matches the lower bound that is calculated equal to $8.5~\mathrm{meV}$ in considering $\tau_{\mathrm{e}}\sim \tau_{\mathrm{tr}}=34~\mathrm{nm}$ in bilayer graphene because of the $2\pi$ Berry's phase and ignoring that $\tau_{\mathrm{tr}}$ could be larger than $\tau_{\mathrm{e}}$ because of the long-range character of the dominant scattering potential. On the other hand, the extrapolation of the linear behavior of $\Delta E_{\mathrm{b}}(\nu)$ at zero energy between $\nu=-4$ and $\nu=0$ leads to $\nu_{\mathrm{edge}(n=0,1)}\simeq -3$, which corresponds to a mobility edge shifted from the Landau level center by a larger energy of $34.5~\textrm{meV}$ ($400~\textrm{K}$). This value cannot be explained by the model of the broadening by disorder valid for $n=-2$ but could be related to the degeneracy of the n=0 and n=1 LLs.

\subsection{\label{sec:level2} Longitudinal resistance reproducible fluctuations}

Fig.~\ref{fig3}c shows that in the BL sample, at both $T=0.35~\textrm{K}$ and $T=1.5~\textrm{K}$, the pattern of reproducible fluctuations of \rx~in the low current regime at $I=0.5~\mu\mathrm{A}$ is similar to that measured at  $I=1~\mu\mathrm{A}$, where conductivity mainly results from the QUILLS mechanism. Measurements at currents above the breakdown current ($\geqslant 2~\mu\mathrm{A}$) have shown a strong decrease of the relative amplitude of these fluctuations. From these observations we deduce that the QUILLS mechanism adds a conductivity contribution that does not itself fluctuate with carrier density. Only the term $\sigma_{T,\nu}$ manifesting itself in the low current regime IV has fluctuations. The resistance shift due to the QUILLS mechanism by increasing current from $0.5~\mu\mathrm{A}$  up to $1~\mu\mathrm{A}$ is particularly visible in the inset of Fig.~\ref{fig3}c which reports $\ln(R_{\mathrm{xx}})$ as a function of $n_{\mathrm{s}}$ between $-1.9\times 10^{12}~\textrm{cm}^{-2}$ and $-2.3\times 10^{12}~\textrm{cm}^{-2}$ away from the \rx~minimum at $n_{\mathrm{s}}=-1.8\times 10^{12}~\textrm{cm}^{-2}$. It also shows a quite linear relationship between $\ln(R_{\mathrm{xx}})$ and $n_{\mathrm{s}}$ for both currents. At $I=1~\mu\mathrm{A}$ such a behavior is expected since this is a feature of the QUILLS mechanism, as also observed at higher current in Fig.~\ref{fig9}a and Fig.~\ref{fig9}b. On the other hand, at $I=0.5~\mu\mathrm{A}$ in current regime IV, QUILLS cannot account for the linear behavior since it is no more the dominant dissipation mechanism, as observed in current dependence of conductivity in Fig.~\ref{fig7}. But it turns out that VRH with soft Coulomb gap predicts $\ln(R_{\mathrm{xx}})\propto -(T_0(\xi_{\mathrm{loc}})/T)^{1/2} \propto -\xi_{\mathrm{loc}}^{-1/2}$. Assuming $\xi_{\mathrm{loc}}\propto (\nu-\nu_{\mathrm{c}})^{-2.3}$, VRH also leads to a near linear behavior of logarithmic conductivity with $\nu$, since $\ln(R_{\mathrm{xx}})$ should be proportional to $(\nu-\nu_{\mathrm{c}})^{2.3/2}$. At $\nu=-4.8$ ($n_{\mathrm{s}}=-2.2\times 10^{-12}~\mathrm{cm^{-2}}$), VRH would lead to $T_0=5~\mathrm{K}$ and $\xi_{\mathrm{loc}}=800~\mathrm{nm}$. VRH can also explain that fluctuation amplitude decreases as $\nu$ increases (in absolute value), and that it decreases slightly with increasing temperature and more strongly with increasing current. This mechanism indeed predicts longitudinal resistance fluctuations resulting from gaussian fluctuations of the localization length $\xi_{\mathrm{loc}}$ with an amplitude $\delta\ln(R_{\mathrm{xx}})\propto T^{(-1/2)}\xi_{\mathrm{loc}}^{(-3/2)}\delta\xi_{\mathrm{loc}}$ that decreases as temperature and $\xi_{\mathrm{loc}}$ increase. Decreasing of the amplitude with current can be explained by the heating effect by current becoming very significant far from $\nu=-4$ for currents near $1~\mu\mathrm{A}$, as observed in regime III for $T=0.35~\textrm{K}$ when $\nu$ decreases from $-4.2$ to $-4.8$ (Fig.\ref{fig7}): $T^{\ast}(1~\mathrm{\mu A})=0.5~\mathrm{K}$ at $\nu=-4.2$ increases up to, $T^{\ast}(1~\mathrm{\mu A})=5.5~\mathrm{K}$ at $\nu=-4.8$. Thus, the reproducible fluctuations of \rx~observed are compatible with the existence of VRH in the regime IV at low current and low temperature. Coulomb blockade in compressible islands surrounded by incompressible strips could also be considered as a source of conductivity fluctuations. In this hypothesis, peaks of conductance would correspond to the addition of one electron into islands\cite{Cobden1999} and $\pi\Delta n_{\mathrm{s}} {{r}^2}=1$ with $\Delta n_{\mathrm{s}}$ the carrier density width of peaks. Considering the experimental value of $\Delta n_{\mathrm{s}}$ leads to typical radius of islands $r$ between $43~\mathrm{nm}$ to $70~\mathrm{nm}$, thus in agreement with values found by others groups\cite{Branchaud2010,Martin2009,Jung2011} but also not so far from the puddle correlation length at low field (11 nm). Although it is difficult to conclude about the mechanism at the origin of fluctuations, they can explain fluctuations of the Hall resistance \rh~observable in Fig.~\ref{fig4}a due to the unavoidable residual coupling between \rh~and \rx.

\section{\label{sec:level2} Conclusion}

To conclude, we have performed quantization tests of the QHE in $\mu \mathrm{m}$ wide Hall bars based on bilayer and monolayer exfoliated graphene deposited on $\textrm{Si/SiO}_{2}$ substrate where electronic transport properties at low magnetic field are mainly governed by the Coulomb interaction of carriers with a high concentration of charged impurities. On the Hall plateaus corresponding to Landau level filling factor near $\nu=2$ in the ML sample and $\nu=-4$ in the BL sample, the Hall resistance \rh~respectively agrees with \rk$/2$ and \rk$/4$ within a relative uncertainty of a few parts in $10^{7}$, in the limit of zero dissipation or at low current below a few $\mu\mathrm{A}$. These experiments are therefore the most accurate QHE quantization measurements to date in monolayer and bilayer exfoliated graphene. They contribute to generalize the universality property of $R_\mathrm{K}$ to the bilayer graphene material for which the QHE was not investigated metrologically so far. At low magnetic field, charged impurities probably located in the silicon substrate at about $1~\textrm{nm}$ below the surface and with density near $2\times 10^{12}~\textrm{cm}^{-2}$ reduce mobility, more strongly in the BL sample ($\mu<2300~\mathrm{cm^{2}\mathrm{V}^{-1}\mathrm{s}^{-1}}$) than in the ML sample ($\mu<4050~\mathrm{cm^{2}\mathrm{V}^{-1}\mathrm{s}^{-1}}$). These very efficient long-range scatterers also induce large spatial fluctuations of carrier density that stays bipolar up to finite density values ($2\times 10^{12}~\textrm{cm}^{-2}$ in BL). Such density inhomogeneity can notably be responsible for the saturation of $L_\Phi$ observed in the BL sample at low temperature and at finite density. In the QHE regime, dissipation leading to the QHE breakdown mainly occurs through quasi-elastic inter-Landau level scattering (QUILLS) in presence of high local electric fields. We claim that a high concentration of charged impurities very close to graphene efficiently assist elastic inter-Landau levels transitions. In addition, charged impurities induce a strong filling factor spatial inhomogeneity which is favorable to the existence of large local electric fields. At low temperature and low current, it is observed in the BL sample that dissipation also follows an activation law with a typical energy of $95~\mathrm{\mu eV}$, the origin of which is not understood. As a result, breakdown is very anticipated at currents as low as $1~\mathrm{\mu A}$ by enhancement of the inter-Landau level transitions which prevent from measuring the Hall resistance quantization with better accuracy at higher currents. This is even more tragic in the small graphene samples produced by exfoliation technique. The role of charged impurities present in the ML sample is expected to be qualitatively the same in the anticipated breakdown, but possibly with quantitative differences resulting from particularities of the Coulomb potential screening.
We then conclude that the development of a graphene based quantum resistance standard able to challenge GaAs would require large samples with higher mobility and more homogeneous carrier density. To achieve this, the role of substrate on which graphene is deposited or grown has to be carefully addressed whatever the graphene fabrication technique considered. This is a consequence of the high sensitivity of graphene electronic transport properties to its environment.

\begin{acknowledgments}
We wish to acknowledge K. Bennaceur for advices in sample fabrication, C. Chaubet and J.-N. Fuchs for valuable discussions.
\end{acknowledgments}

\providecommand{\noopsort}[1]{}\providecommand{\singleletter}[1]{#1}%

\end{document}